\documentclass[aps,prl,twocolumn,superscriptaddress]{revtex4-1}

\usepackage{mathrsfs,amsmath,amssymb,amsthm,natbib}
\usepackage{amssymb,fmtcount}
\usepackage{graphicx,epsfig,latexsym,overpic,amssymb,color}
\usepackage{threeparttable}

\usepackage{url}
\usepackage[colorlinks,
linkcolor=blue,
anchorcolor=blue,
urlcolor=blue,
citecolor=blue]{hyperref}
\begin{document}


\title{Spatial Entanglement Entropy in Nuclear Fission}
\author{S.C. Li}\thanks{These authors contributed equally to this work.}
\affiliation{
School of Physics, State Key Laboratory of Nuclear Physics and Technology,
Peking University, Beijing 100871, China
}
\author{J.W. Chen}\thanks{These authors contributed equally to this work.}
\affiliation{
School of Physics, State Key Laboratory of Nuclear Physics and Technology,
Peking University, Beijing 100871, China
}
\author{J.C. Pei}\email[Corresponding author:]{peij@pku.edu.cn}
\affiliation{
School of Physics, State Key Laboratory of Nuclear Physics and Technology,
Peking University, Beijing 100871, China
}
\affiliation{
Southern Center for Nuclear-Science Theory (SCNT), Institute of Modern Physics, Chinese Academy of Sciences, Huizhou 516000,  China
}

\date{\today}

\begin{abstract}
Nuclear fission provides a unique manifestation of spatially nonlocal many-body entanglement.
We compute the bipartite spatial entanglement entropy exactly along dynamical fission trajectories,
by leveraging the fermionic Gaussian state formulation.
Across seven representative fissioning channels, the final entanglement entropy correlates strongly with the intrinsic particle number variance of the fragment, yet exhibits no simple dependence on scission geometries.
Most notably the entanglement is significantly suppressed when fragments are magic nuclei, revealing a shell anti-entanglement effect. 
This work establishes entanglement entropy as a novel lens 
 that extends the conventional conception of nuclear fission.
\end{abstract}

\maketitle

\textit{Introduction.---}
Entanglement entropy, a universal information-theoretic metric,
quantifies  the degree of entanglement between subsystems and
serves as a powerful probe of exotic phases and phase transitions~\cite{AmicoEtAl2008}
, underlying symmetries~\cite{BeaneEtAl2019,ares}
, and even the emergence of spacetime geometry~\cite{BoseEtAl2017,Swingle2018}.
Nuclear fission stands out as a unique and naturally occurring manifestation of spatially nonlocal many-body entanglement in strongly interacting systems.
The final stage of fission is a rapid non-adiabatic scission  process in a few 10$^{-22}$
seconds; consequently, substantial wave-function nonlocality persists after two fragments fly apart~\cite{QiangPeiGodbey2025}.
This dynamical entanglement alters the partition of particles and energy between two well-separated fragments~\cite{QiangPeiGodbey2025,ShangQiangPei2025}, challenging the widely accepted concept
that particle exchange occurs exclusively through the neck~\cite{BrosaGrossmannMuller1990,bulgac2020}.
To characterize nuclear fission via the entanglement lens,
the primary task is therefore to  compute the entanglement entropy accurately.

The bipartite entanglement entropy is defined by the reduced density matrix of either subsystem and can in principle be computed via the Schmidt decomposition.
For many-body systems, the direct Schmidt decomposition is impossible due to its exponential growth in Hilbert space.
For a Slater determinant wave function from Hartree-Fock calculations, the entanglement entropy between two spatially separated subsystems can
be calculated exactly~\cite{Klich2006}.
For nuclear fission, pairing correlations are essential for accelerating both static fission tunneling~\cite{SadhukhanEtAl2014} and dynamical evolutions~\cite{TanimuraLacroixScamps2015,ScampsSimenelLacroix2015,Bulgac2016}.
It is nontrivial to calculate entanglement entropy based on the non-adiabatic time-dependent density functional approach (TDDFT) with pairing correlations,
and earlier calculations have not been strictly validated~\cite{QiangPeiGodbey2025}.
On the other hand, fermions with pairing correlations can be described by the formulation of Gaussian state~\cite{PeschelEisler2009}, which
enables exact calculations of reduced density matrix and entanglement entropy~\cite{FagottiCalabrese2008,PuspusVillegasParaan2014,SahaKulkarniDhar2024}.
This method has wide applications in condensate matter but has not been applied in nuclear systems yet.

Nuclear fission is conventionally classified by two geometric features of the scission configuration: the mass asymmetry, which varies from symmetric to asymmetric and super-asymmetric channels~\cite{BrosaGrossmannMuller1990,PoenaruEtAl2012,WardaZdebRobledo2018}, and the neck length, which ranges from super-long to super-short modes~\cite{BrosaGrossmannMuller1990,AlbertssonEtAl2021}.
Furthermore, the shell effects in nascent fragments are important to explain the systematics of fission yields~\cite{itkis,ScampsSimenel2018,MorfouaceEtAl2025}.
The experiment on $^{180}$Hg reveals that the fission pathway is also crucial~\cite{AndreyevEtAl2010}.
The microscopic TDDFT provides consistent descriptions of non-equilibrium non-adiabatic fission dynamics, pre-scission shapes,
 and the partition of particles, energies, and angular momentum between binary fragments, although collective
 shape fluctuations are insufficient in TDDFT~\cite{SimenelUmar2014,stevenson,Nakatsukasa2016,Bulgac2016,godbey,luguo,Simenel2025,QiangPeiGodbey2025}.
 Generally the geometric boundary  is closely related to the scaling laws of entanglement entropy, suggesting
 the scission geometry may also be relevant.
 While the area law scaling is ubiquitous in conventional many-body systems~\cite{EisertCramerPlenio2010},  nuclear systems typically exhibit volume law scaling
due to complex correlations~\cite{GuSunHagenPapenbrock2023}.
 It is of great interest to investigate how the scission geometry, shell effects, and dynamical evolutions influence the entanglement entropy in fission,
thereby providing new opportunities to explore many-body entanglement.

In this Letter, we present an accurate calculation of entanglement entropy in nuclear fission along the time-dependent Skyrme
Hatree-Fock+BCS  trajectories~\cite{QiangPeiStevenson2021}, leveraging the fermionic Gaussian state formulation~\cite{PeschelEisler2009}.
This method was originally derived in the configuration space and
has not been applied to spatially separated subsystem before.
For cross-validation, we implement two independent approaches within this framework: the Grassmann kernel method~\cite{ChungPeschel2001} and the covariance matrix method~\cite{Peschel2003,PeschelEisler2009},
to calculate the entanglement entropy .
In nuclear physics, entanglement entropy in static configuration space has been studied extensively~\cite{RobinSavagePillet2021,KruppaEtAl2022,BaiRen2022,JafarizadehEtAl2022,GuSunHagenPapenbrock2023,PerezObiolEtAl2023,JohnsonGorton2023,ChenFrauendorf2024,XuEtAl2026},
and these studies have motivated optimized truncation
schemes for many-body calculations~\cite{GortonJohnson2024}.
The distributions of fission observables are conventionally described by the random neck rupture with different scission geometries ~\cite{BrosaGrossmannMuller1990,randrup2011}.
Within this picture,  wide distributions of fission yields correspond to elongated scission necks.
Hereby the double particle number  projection
in spatial subspace is employed to obtain the post-scission distributions of fission yields owing to intrinsic fluctuations~\cite{AnguianoEgidoRobledo2001,Simenel2010,VerriereSchunckKawano2019}.
Seven selected fission systems span conventional asymmetric actinide fission ($^{236}$U and $^{240}$Pu),
the evolution from asymmetric to shell-driven symmetric splitting along the fermium chain ($^{256,258,264}$Fm)~\cite{HoffmanLane1995,lay},
and $^{208}$Pb accompanied cluster-like or super-asymmetric channels ($^{222}$Ra~\cite{zhang2026} and  $^{274}$Hs~\cite{Yanez2014,WardaZdebRobledo2018}).
This set allows the effects of  scission geometry and fragment shell structures to be
contrasted by the universal entanglement metric.

\textit{Methods.---}
The
fissioning nucleus is divided into two complementary spatial regions $A$ and $B$,
separated in the neck at $z=z_{\rm neck}$~\cite{Klich2006}.
The one body projectors are $P_A$, $P_B=I-P_A$, with
$P_A\psi(\mathbf r)=\theta_A(\mathbf r)\psi(\mathbf r)$, where
$\theta_A$ is the indicator function of region~$A$.
By constructing the overlap matrix $M^A_{ij}=\langle\phi_i|P_A|\phi_j\rangle$
and diagonalizing the matrix $U^\dagger M^A U=d$,
we obtain eigenvalues
$0\leq d_\alpha\leq1$ that quantify the spatial distribution of each
mode.
Modes with $0<d_\alpha<1$ are split into localized orbitals
according to
\begin{equation}
\begin{aligned}
\widetilde c_\alpha^\dagger&= \sqrt{d_\alpha}\,a_\alpha^\dagger+\sqrt{1-d_\alpha}\,b_\alpha^\dagger,
\qquad \widetilde c_\alpha^\dagger= \sum_i U_{i\alpha}c_i^\dagger,
\\
\widetilde e_\alpha^\dagger&= \sqrt{1-d_\alpha}\,a_\alpha^\dagger-\sqrt{d_\alpha}\,b_\alpha^\dagger,
\end{aligned}
\label{eq:mode-splitting}
\end{equation}
where $c_i^\dagger$ corresponds to Hartree-Fock states, $\widetilde e_\alpha^\dagger$ denotes auxiliary operator introduced for a complete unitary transformation from full space to subpsace basis, and $a_\alpha^\dagger$, $b_\alpha^\dagger$ create particles
localised in $A$ and $B$ subspaces, respectively.
With pairing correlations, the quasiparticle state can be reformulated into a fermionic Gaussian state~\cite{PeschelEisler2009}.
The Gaussian state representation provides the natural language for the reduced density matrix,
facilitating the calculation of entanglement entropy~\cite{Peschel2003}.
Within this framework, we employed two approaches:  Grassmann kernel method
~\cite{ChungPeschel2001} and the covariance matrix method~\cite{Peschel2003,PeschelEisler2009}, for exact calculations of entanglement entropy.

\textit{Grassmann kernel method:} The BCS ground state can be written in the exponential form $|\Phi_0\rangle = C_1 \exp\!\Bigl(\sum_{i=1}^n \frac{v_i}{u_i}\,
c_i^\dagger c_{\bar i}^\dagger\Bigr)|0\rangle$, where  $C_1 = \prod_{i=1}^n u_i$.
By substituting the spatial decomposition with the localized operators
from Eq.~\eqref{eq:mode-splitting}
into this exponential form and then grouping terms by their spatial
label ($A$ or $B$), we get
\begin{equation}
|\Phi_0\rangle = C_1 \exp\!\left(\frac12\sum_{ij} G_{ij}\,
e_i^\dagger e_j^\dagger\right)|0\rangle,
\label{eq:G-def}
\end{equation}
where $e^\dagger=[a_1^\dagger,\dots,a_{\bar n}^\dagger,b_1^\dagger,\dots,b_{\bar n}^\dagger]$.

The reduced density matrix $\rho_A=\mathrm{Tr}_B|\Phi_0\rangle\langle\Phi_0|$
can be evaluated in the fermionic coherent state representation, following the reduced density matrix construction for free fermions~\cite{ChungPeschel2001,Peschel2003,PeschelEisler2009} and the Grassmann representation of fermionic Gaussian maps~\cite{Bravyi2005}.
The partial trace reduces to a Grassmann Gaussian integral over the
$B$ variables. The Grassmann Gaussian integral results in
\begin{equation}
\begin{aligned}
\rho_A
=&C_2\exp\!\left(\sum_{ij}\frac{1}{2}\alpha_{ij}\xi_i^*\xi_j^*\right)\exp\!\left(\sum_{ij}(\ln\beta)_{ij}\xi_i^*\xi_j^\prime\right)*\\
& \exp\!\left(\sum_{ij}\frac{1}{2}\gamma_{ij}\xi_{i}^{\prime}\xi_{j}^{\prime}\right),\quad i,j\leqslant M.\\
\end{aligned}
\label{eq:rhoA-factorized}
\end{equation}
where the matrices $\alpha,\beta,\gamma$ are expressed in terms of
the blocks of $G$, and $C_2$ is a normalization constant.

The factorised density matrix~\eqref{eq:rhoA-factorized} is brought
back to the single exponential Gaussian form through the inverse
Balian--Br\'ezin decomposition~\cite{BalianBrezin1969}:
\begin{equation}
    \rho_A \propto  \exp\!\left(\frac{1}{2}\xi^\dagger K\xi\right)
\end{equation}
where $\xi^\dagger=(a^\dagger, a)=(A_1^\dagger,\dots,A_{\bar{n}}^\dagger,A_1,\dots,A_{\bar{n}})$,
and the generating matrix $e^K$ is constructed from $\alpha,\beta,\gamma$,
\begin{equation}
\begin{aligned}
e^{K}&=
\begin{pmatrix}
\beta+\alpha\beta^{-T}\gamma & \alpha\beta^{-T}\\\beta^{-T}\gamma&\beta^{-T}
\end{pmatrix}.
\end{aligned}
\end{equation}
Diagonalising $K$ gives the eigenvalue spectrum  $\{\varepsilon_l\}$.
The von Neumann entropy $S_E$ is given in terms of the occupation numbers
$n_l = 1/(1+e^{\varepsilon_l})$ as:
\begin{equation}
S_E=S_A = \sum_{l=1}^{2n}\bigl[-(1-n_l)\ln(1-n_l)-n_l\ln n_l\bigr].
\label{eq:S-wavefunction-entropy}
\end{equation}

\begin{figure*}[!htbp]
    \centering
    \includegraphics[width=\linewidth]{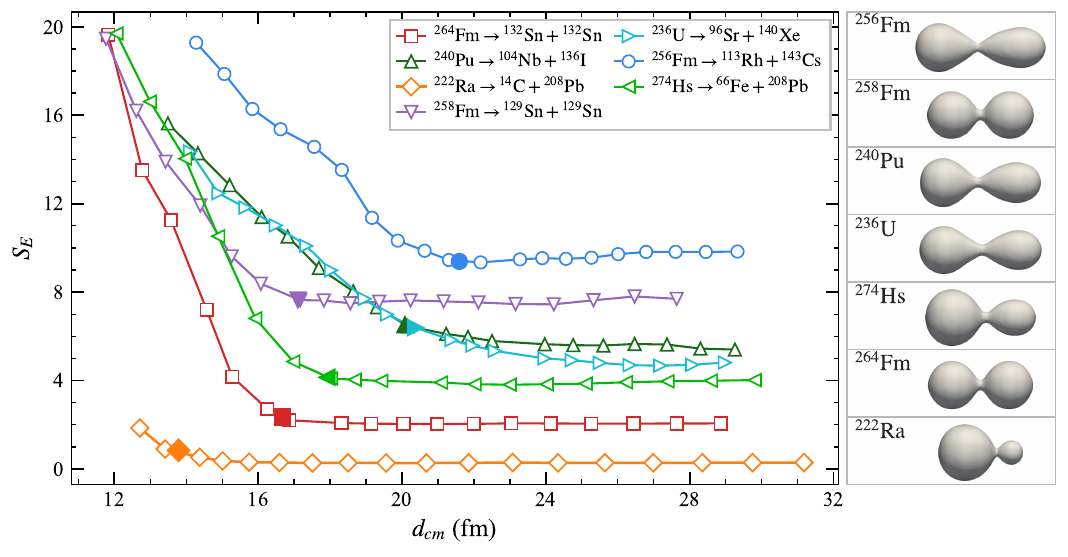}
    \caption{Evolution of entanglement entropy $S_E$ between two fragments as a function of their distances $d_{cm}$ for different fissioning systems.
    The solid symbols denote the scission configuration at the particle number in neck $N_c$=0.1. The right panel shows the scission shapes at the same criteria of  $N_c$=0.4.
    }
    \label{fig:entropy}
\end{figure*}

\textit{Covariance matrix method:}
The Majorana covariance matrix, as a standard representation of fermionic Gaussian states,  provides an alternative method to calculate entanglement entropy in superconducting systems~\cite{ChungPeschel2001,BravyiKitaev2002,KrausWolfCiracGiedke2009,PeschelEisler2009}.
Introduce Majorana operators
$x_i=c_i+c_i^\dagger$, $y_i=-i(c_i-c_i^\dagger)$
and the block ordered vector
$w=(x_1,\dots,x_{2n},y_1,\dots,y_{2n})^T$.
The Majorana covariance matrix is defined as
$\Gamma_{lm}= \frac{i}{2}\mathrm{Tr}\bigl(\hat\varrho\,[w_l,w_m]\bigr)$,
which is real, antisymmetric and can also be connected to the generalized density matrix in the Bogoliubov transformation.

The covariance matrix is transformed into $\widetilde\Gamma = O_U\Gamma O_U^T$ by using $\widetilde c_\alpha^\dagger$ operators in Eq.(\ref{eq:mode-splitting}),
where $O_U$ is the orthogonal transformation induced by the unitary $U$.
After some derivation, we obtain the covariance matrices of the two subspace regions as
\begin{equation}
\begin{aligned}
\Gamma_A &= \mathbb D_A\widetilde\Gamma\mathbb D_A
          + \mathbb D_B\Gamma_0\mathbb D_B,
\\
\Gamma_B &= \mathbb D_B\widetilde\Gamma\mathbb D_B
          + \mathbb D_A\Gamma_0\mathbb D_A,
\\
C_{AB}   &= \mathbb D_A\widetilde\Gamma\mathbb D_B
          - \mathbb D_B\Gamma_0\mathbb D_A.
\end{aligned}
\end{equation}
where $\mathbb D_A$, $\mathbb D_B$ are diagonal matrix built from $\sqrt{d_\alpha}$ and $\sqrt{1-d_\alpha}$, respectively.
The auxiliary operators in Eq.(\ref{eq:mode-splitting}) corresponds to particle vacuum states, and $\Gamma_0$ denotes the auxiliary vacuum covariance matrix.
Assembling these blocks gives the full bipartite covariance matrix
\begin{equation}
\Gamma^{(AB)} =
\begin{pmatrix}
\Gamma_A  & C_{AB}\\
-C_{AB}^T & \Gamma_B
\end{pmatrix}.
\label{eq:full-spatial-covariance}
\end{equation}

The reduced density operator $\hat\varrho_A=\mathrm{Tr}_B\hat\varrho$
remains a fermionic Gaussian state, based on the Grassmann Gaussian integral~\cite{KrausWolfCiracGiedke2009,SzalayEtAl2021}.
Consequently, the covariance matrix of the reduced state is simply the
principal block of $\Gamma^{(AB)}$ corresponding to region~$A$:
\begin{equation}
\Gamma_A^{\rm red}= \Gamma^{(AB)}\big|_A = \Gamma_A.
\label{eq:restricted-covariance}
\end{equation}
Thus the exponential complexity of the partial trace is avoided, by invoking only
the single particle covariance spectrum.

The real antisymmetric matrix $\Gamma_A$ can be rewritten as a canonical form
by an orthogonal transformation,
\begin{equation}
O_A\Gamma_AO_A^T = \bigoplus_\mu
\big(\begin{pmatrix}0&-\nu_\mu\\ \nu_\mu&0\end{pmatrix}\big)
,\qquad 0\leq\nu_\mu\leq 1
\end{equation}
The von Neumann entropy $S_E$ is obtained from the spectrum $\{\nu_\mu\}$, as in superconducting entanglement spectrum analyses~\cite{Peschel2003,PeschelEisler2009,DiTullioGigenaRossignoli2018}:
\begin{equation}
S_E= -\sum_{\mu=1}^{2n}
\Bigl[ \frac{1+\nu_\mu}{2}\ln\frac{1+\nu_\mu}{2}
      +\frac{1-\nu_\mu}{2}\ln\frac{1-\nu_\mu}{2} \Bigr],
\label{eq:covariance-entropy}
\end{equation}
 The numerical results are validated by exact diagonalization  for small
systems ($n\leq3$). We demonstrated that the two approaches deliver exactly the same results. Further details of the theoretical methods are provided
in the Supplemental Material~\cite{supp}.

\begin{figure*}[!htbp]
    \centering
    \includegraphics[width=\linewidth]{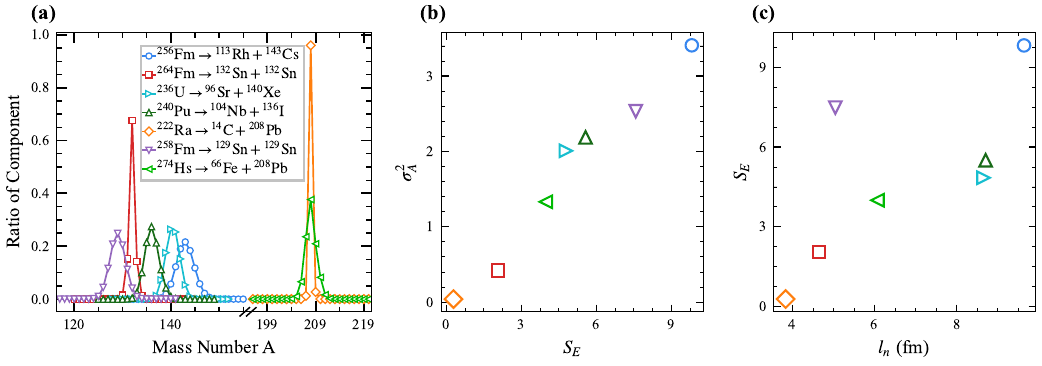}
    \caption{(a) The calculated distributions of heavy fragment mass yields for different fissioning systems.
    (b) the particle number variances $\sigma_A^2$ against the final entanglement entropy $S_E$.
    (c) the scission neck length $l_n$ versus $S_E^{}$.}
    \label{fig:ccb}
\end{figure*}

\textit{Results.---}
Seven representative fissioning systems are investigated as shown in Fig.~\ref{fig:entropy}.
The entanglement entropy  $S_E$  is shown as a function of the center-of-mass distance $d_{cm}$  between two fragments.
The $S_E$ decreases along the scission process, in which neck contraction
and  nucleon reorganization drives the localization of wave functions. Finally substantial
nonlocality survives due to the fast violent splitting and $S_E$  becomes nearly stationary at large distances~\cite{QiangPeiGodbey2025}.
The quasi-plateau behavior of $S_E$  at large $d_{\rm cm}$ indicates the freezing of interfragment quantum correlations.
In  Fig.~\ref{fig:entropy},  the number of particles in the neck~\cite{QiangPeiGodbey2025} $N_c$=0.1 as an indication of scission corresponds to the entanglement freezing, although
the partition of average number of particles was thought to be much earlier~\cite{RegnierEtAl2016,bulgac2020}.

The most striking finding in Fig.~\ref{fig:entropy} is that the entanglement entropy is strongly suppressed
when fragments are magic nuclei. For $^{264}$Fm splitting into double $^{132}$Sn, $S_E$
is much smaller than that of $^{258}$Fm, although both have the same short-neck configuration.
For $^{222}$Ra with $^{14}$C emission~\cite{zhang2026}, the entanglement is close to zero, demonstrating that cluster emission
is distinctly different from fission. The entanglement of super-asymmetric fission of $^{274}$Hs with
the $^{208}$Pb  fragment is also suppressed. The significant suppression of entanglement when fission
fragments are magic nuclei demonstrated the shell anti-entanglement effect, irrespective of symmetric or asymmetric fission modes.
This can be explained that the fragment shell structures favors the localization of wave functions~\cite{ZhangEtAl2016}.
Recently the entanglement suppression was proposed to be the potential origin of emergent symmetries in hadron scatterings~\cite{BeaneEtAl2019,HuEtAl2025}.
This can be understood similarly that magic fragments stabilize symmetries and suppress the entanglement.
The final entanglement entropy forms a constant plateau with magic fragments to resist dynamical fluctuations, while other systems
exhibit small variations.
It has to be noted that deformed shells such as octupole deformation shells are also important in nascent fragments of
actinide nuclei, although such effects are relatively less pronounced compared to magic fragments~\cite{ScampsSimenel2018,HanWardaZdebRobledo2021}.
From this respective, the decreasing entanglement in fission is generally  attributed to the shell-driven localization, which competes with the survival of entanglement owing to rapid non-adiabatic dynamics.

Figure~\ref{fig:ccb} shows the calculated distributions of fission yields and their relationship to $S_E$ for different fissioning systems.
The distributions of fission mass yields are obtained by the double particle number projection in spatial subspaces, which measures the components of the entangled state with different fragment particle
numbers.
In Fig.~\ref{fig:ccb}(a),  the mass distribution widths around magic fragments are very narrow and the distribution of heavy fragments from $^{274}$Hs is relatively narrow.
The widths of other systems are more or less similar and are around 4 mass units.
Note that the present double projection calculations employ the recent corrected formula~\cite{Robledo2026}.
The calculated distribution widths associated with intrinsic fluctuations  are not enough compared to experimental widths, because either high-order correlations
or collective fluctuations are absent beyond the TDDFT framework~\cite{LacroixAyik2014,Simenel2025,BenderEtAl2020}.

The particle number variance $\sigma_A^2$ of fission yields in terms of the final entanglement entropy is shown in Fig.~\ref{fig:ccb}(b).
The results demonstrate a strong positive correlation between $\sigma_A^2$ and $S_E^{}$: systems with larger post scission $S_E^{}$ generally exhibit larger particle number variances.
This observation is consistent with the broader relation between entanglement entropy and  intrinsic particle number fluctuations in fermionic systems~\cite{KlichLevitov2009,SongRachelLeHur2010,CalabreseMintchevVicari2012,PuspusVillegasParaan2014}.
However, this correlation is not exactly a linear relation between $S_E$ and $\sigma_A^2$, considering the influences of specific fission modes and shell structures.
Fig.~\ref{fig:ccb}(c) shows the relationship between geometric neck lengths  and $S_E^{}$.
The scission neck length $l_n$ is defined as $l_n=d_{\rm cm}-r_0\left(A_L^{1/3}+A_R^{1/3}\right)$ at the same criterion $N_c$=0.1, in which $r_0=1.2~\mathrm{fm}$, $A_L$ and $A_R$ are fragment mass numbers.
There is no simple monotonic relationship between the neck length and the post scission $S_E$.
This indicates that the geometric neck length near scission  cannot determine the final interfragment entanglement alone.

\begin{figure}[!htbp]
    \centering
    \includegraphics[width=\linewidth]{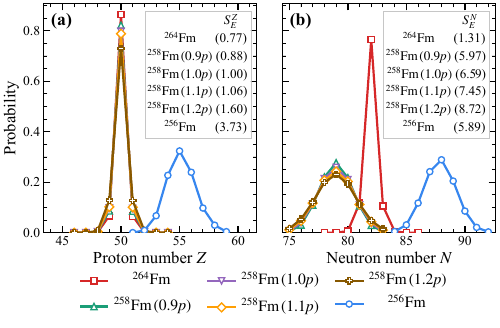}
    \caption{ Distributions of Fission yields as a function of neutron and proton numbers in $^{256,258,264}$Fm isotopes.
    For $^{258}$Fm, calculations with a varying scale of pairing strength from 0.9 to 1.2  are compared.
    The corresponding entanglement entropy of protons $S_E^Z$ and neutrons  $S_E^N$ are given in the bracket. }
    \label{fig:fm}
\end{figure}

A noteworthy example is $^{258}\mathrm{Fm}$, for which the neck length is relatively short,  while its post scission $S_E$ is large.
In Fig.\ref{fig:fm}, the detailed results actually show $S_E^{}$ of protons in $^{258}$Fm is close to $^{264}$Fm, and its width of fission yields is dominated by neutron entanglement that is close to $^{256}$Fm.
This again supports the suppression of entanglement by shell effects.
Distributions of experimental mass yields of $^{258}$Fm is rather narrow~\cite{HoffmanEtAl1980}, indicating that there is little room for collective shape fluctuations for this short neck channel. Charge yields of $^{258}$Fm are expected to be even narrower and future measurements are needed.
For $^{258}$Fm with a varying scale of pairing strength from 0.9 to 1.2, the $S_E^N$ of neutrons
increases from 5.97 to 8.72. The widths also increase, with the neutron number variance $\sigma_N^2$ changing from
2.00 to 2.98.  Entanglement entropy is
known to increase with pairing correlations in superconducting
systems~\cite{PuspusVillegasParaan2014,OliveiraSacramento2014}.
As the pairing strength increases, $S_E^N$ increases by 46$\%$, scaling well with the 49$\%$ increase in $\sigma_N^2$ .
The trend is similar for protons although the entanglement is weaker.
This work focuses on the relation among widths of fission yields, entanglement entropy, and scission geometries.
Entanglement is also expected to be relevant to
other fission observables such as the energy partition and fragment spins~\cite{bulgac2022,scamps,ShangQiangPei2025,Qiang2026}.

\textit{Summary.---}
In this Letter, we formulate the entanglement
entropy in nuclear fission using the fermionic
Gaussian-state representation to characterize spatially
nonlocal many-body entanglement and the scission mechanism.
Two formulations including the Grassmann kernel method and the covariance matrix method are implemented for cross-validation.
The entanglement entropy of seven selected fissioning systems along TDDFT dynamical trajectories are studied,
with varying fission modes and scission geometries.
Results reveal strong correlation between the entanglement entropy and the intrinsic particle variances of fragments, irrespective of
the scission-neck lengths.
The most notable finding is that the entanglement is strongly suppressed when fragments are magic nuclei, revealing
a shell anti-entanglement effect.
This suppression arises because shell closure favors wave-function localization and stabilizes  symmetries, offering
profound perspectives on dynamical many-body entanglement.
It is worth expecting that entanglement as a new lens will bring more transformative insights on nuclear fission.

\textit{Acknowledgments---}
We thank useful discussions with Y. Qiang, Y.N.Zhang and F.R. Xu.
 This work was supported by  the
 National Key R$\&$D Program of China (Grant  No.2023YFA1606403),
  the National Natural Science Foundation of China under Grants No.12475118, 12335007.

\end{document}